\documentclass[iop,apjl]{emulateapj}

\usepackage{fontenc}
\usepackage{amssymb}
\usepackage{amsmath}
\usepackage{graphicx}
\usepackage{xspace}
\usepackage{natbib}
\usepackage[usenames]{color}
\usepackage{paralist}
\usepackage{txfonts}

\shorttitle{\threea}
\shortauthors{Marcu et al.}

\begin{document}

\submitted{Accepted October 18, 2011}

\title{The 5\,hr pulse period and broadband spectrum of the Symbiotic X-ray Binary 3A\,1954$+$319}

\shorttitle{The Symbiotic X-ray Binary 3A\,1954$+$319}

\author{
Diana M.\ Marcu\altaffilmark{1,2}, 
Felix F\"urst\altaffilmark{3},
Katja Pottschmidt\altaffilmark{1,2},
Victoria Grinberg\altaffilmark{3}, 
Sebastian M\"uller\altaffilmark{3},
J\"orn Wilms\altaffilmark{3}, 
Konstantin A.\ Postnov\altaffilmark{4}, 
\mbox{Robin H.\ D.\ Corbet}\altaffilmark{5,2}, 
Craig B.\ Markwardt\altaffilmark{5}, 
and Marion Cadolle Bel\altaffilmark{6}
}

\altaffiltext{1}{NASA Goddard Space Flight Center,
  Astrophysics Science Division, Code 661, Greenbelt, MD 20771, USA}

\altaffiltext{2}{CRESST \& University of Maryland Baltimore County,
  1000 Hilltop Circle, Baltimore, MD 21250, USA} 

\altaffiltext{3}{Dr.\ Karl Remeis-Observatory \& ECAP, University of
  Erlangen-Nuremberg, Sternwartstr.\ 7, 96049 Bamberg, Germany}

\altaffiltext{4}{Sternberg Astronomical Institute, 119992, Moscow, Russia}

\altaffiltext{5}{NASA Goddard Space Flight Center,
  Astrophysics Science Division, Code 662, Greenbelt, MD 20771, USA}

\altaffiltext{6}{European Space Agency, European Space Astronomy
  Centre P.O. Box 78, 28691 Villanueva de la Ca\~{n}ada, 28692,
  Madrid, Spain}

\begin{abstract}
  We present an analysis of the highly variable accreting X-ray pulsar
  \object{3A\,1954$+$319} using 2005--2009 monitoring data obtained
  with \textsl{INTEGRAL} and \textsl{Swift}. This considerably extends
  the pulse period history and covers flaring episodes in 2005 and
  2008.  In 2006 the source was identified as one of only a few known
  symbiotic X-ray binaries (SyXBs), i.e., systems composed of a
  neutron star accreting from the inhomogeneous medium around an
  M-giant star. The extremely long pulse period of $\sim$5.3\,hr is
  directly visible in the 2008 \textsl{INTEGRAL}-ISGRI outburst light
  curve. The pulse profile is double peaked and generally not
  significantly energy dependent although there is an indication of
  possible softening during the main pulse. During the outburst a
  strong spin-up of \mbox{$-1.8\times10^{-4}$\,hr\,hr$^{-1}$}
  occurred. Between 2005 and 2008 a long-term spin-down trend of
  \mbox{$2.1\times10^{-5}$\,hr\,hr$^{-1}$} was observed for the first
  time for this source. The 3--80\,keV pulse peak spectrum of
  3A\,1954$+$319 during the 2008 flare could be well described by a
  thermal Comptonization model. We interpret the results within the
  framework of a recently developed quasi-spherical accretion model
  for SyXBs.
\end{abstract}

\keywords{binaries: symbiotic --- stars: individual (3A\,1954$+$319) ---
  stars: neutron --- X-rays: binaries} 

\section{Introduction}

The X-ray source 3A\,1954$+$319 was detected in the Cygnus region in
surveys by \textsl{Uhuru}, \textsl{Ariel V}, \textsl{EXOSAT}, and
\textsl{ROSAT} \citep{forman:78a,warwick:81a,warwick:88a,voges:99a}.
Pointed observations with \textsl{EXOSAT} \citep{cook:85a} and
\textsl{Ginga} \citep{tweedy:89a} showed a hard X-ray spectrum as well
as intensity variations by an order of magnitude on timescales of
minutes. This led to the suggestion that the system might be a High
Mass X-ray Binary (HMXB). Only when \citet{masetti:06a} identified the
companion as an M4--M5~III star at a distance of $\lesssim$1.7\,kpc
and \citet{corbet:06a} discovered a $\sim$5\,hr pulse period in early
\textsl{Swift}-BAT data, did it become clear that 3A\,1954$+$319 is a
Symbiotic X-ray Binary (SyXB). SyXBs constitute a small
group\footnote{In addition to 3A\,1954$+$319 a recent list of SyXB
  given by \citet{nespoli:09a} consists of: GX\,1$+$4, 4U\,1700$+$24,
  Scutum X-1, IGR\,J16194$-$2810, 1RXS\,J180431.1$-$273932,
  IGR\,J16358$-$4726, and IGR\,J16393$-$4643. Note that the X-ray mass
  function of IGR\,J16393$-$4643 argues against a LMXB classification,
  though \citep{pearlman:11a}.} of persistent Low Mass X-ray Binaries
(LMXBs) in which a neutron star is orbiting in the inhomogeneous
medium around an M-type giant star\footnote{The classification is
  related to but not identical to that of ``Symbiotic Binaries'',
  systems consisting of a white dwarf and a red giant. The latter were
  named after their optical spectra which show contributions from both
  binary components, different from SyXBs where the neutron star is
  not visible at optical wavelengths.}.

SyXBs typically have wide orbits, e.g., the prototype GX\,1$+$4 has an
orbital period of $\sim$1161\,days \citep{hinkle:06a}. Their X-ray
emission is therefore due to wind accretion, a process not well
investigated for late type donors. Two broadband spectral studies
support the SyXB interpretation for 3A\,1954$+$319:
\citet{mattana:06a} modeled the non-simultaneous \textsl{BeppoSAX} and
\textsl{INTEGRAL} spectrum with a highly absorbed
($N_\mathrm{H}\sim10^{23}\,\mathrm{cm}^{-2}$) cutoff power law with a
photon index of 1.1 and a folding energy of 15\,keV and a weak Fe
K$\alpha$ line. These authors also confirmed the detection of the
$\sim$5\,hr period in BAT and ISGRI data. In a study of archival data
spanning absorbed 2--10\,keV luminosities from
$3.4\times10^{34}\,\mathrm{erg}\,\mathrm{s}^{-1}$ to
$1.8\times10^{35}\,\mathrm{erg}\,\mathrm{s}^{-1}$ \citet{masetti:07a}
confirmed the empirical spectral description for the $>$2\,keV
spectrum and determined a best fit using thermal Comptonization,
modified by complex absorption (ionized plus partially covering
neutral absorption). They also modeled a $<$2\,keV soft excess with a
$\sim$50\,eV hot plasma and interpreted it, together with the two-zone
absorption, as being due to a diffuse, partly ionized cloud of
material around the neutron star.

The $\sim$5\,hr period is the only period known for this system. As
\citet{corbet:06a,corbet:08a} argued (i) the period value itself is
inconsistent with being the orbital period of an M-giant, (ii) the
large period decline of
$(2.6\pm0.2)\times10^{-5}\,\mathrm{hr}\,\mathrm{hr}^{-1}$ observed over
the first year of BAT data cannot be due to orbital Doppler
modulation, and (iii) the period change is also too large to be
supported by a white dwarf accretor. Interpreting the $\sim$5\,hr
period as a neutron star spin period makes 3A\,1954$+$319 the slowest
rotating neutron star in an X-ray binary currently known and one of
the slowest pulsars in general -- with only the 6.67\,hr pulsar in the
supernova remnant RCW 103 showing a larger value \citep{deluca:06a}.
\citet{corbet:08a} also noted that if the neutron star rotated close
to its equilbrium period for disk accretion, as usually assumed in
accreting sources, the long period would imply a neutron star magnetic
field of $\sim$$10^{15}$\,G rather than of $\sim$$10^{12}$\,G. The
latter value is more commonly observed for accreting pulsars and is
consistent with a spin-down origin of the long period
\citep{mattana:06a}. From the \textsl{Swift}-BAT and \textsl{RXTE}-ASM
data available at the time it was not clear whether the strong spin-up
observed was associated with the long flaring episode in 2005 and/or
whether the source might show a spin-down at lower fluxes, as it is
the case for GX\,1$+$4 \citep{corbet:08a}.

In this Letter we considerably extend the pulse period history for
3A\,1954$+$319 and present a timing and spectral analysis of a flaring
episode in 2008 (\citealt{fuerst:11a} presented a preliminary analysis
of these data). Section~\ref{sec:obs} describes the observations and
data reduction. Section~\ref{sec:res} reports the results, including
long-term and high time resolution light curves, energy resolved pulse
profiles, the pulse period evolution, and the broadband spectrum. In
Section~\ref{sec:dis} the results are summarized and discussed.

\section{Observations and Data Reduction}\label{sec:obs}

\begin{figure}
  \includegraphics[width=0.5\textwidth]{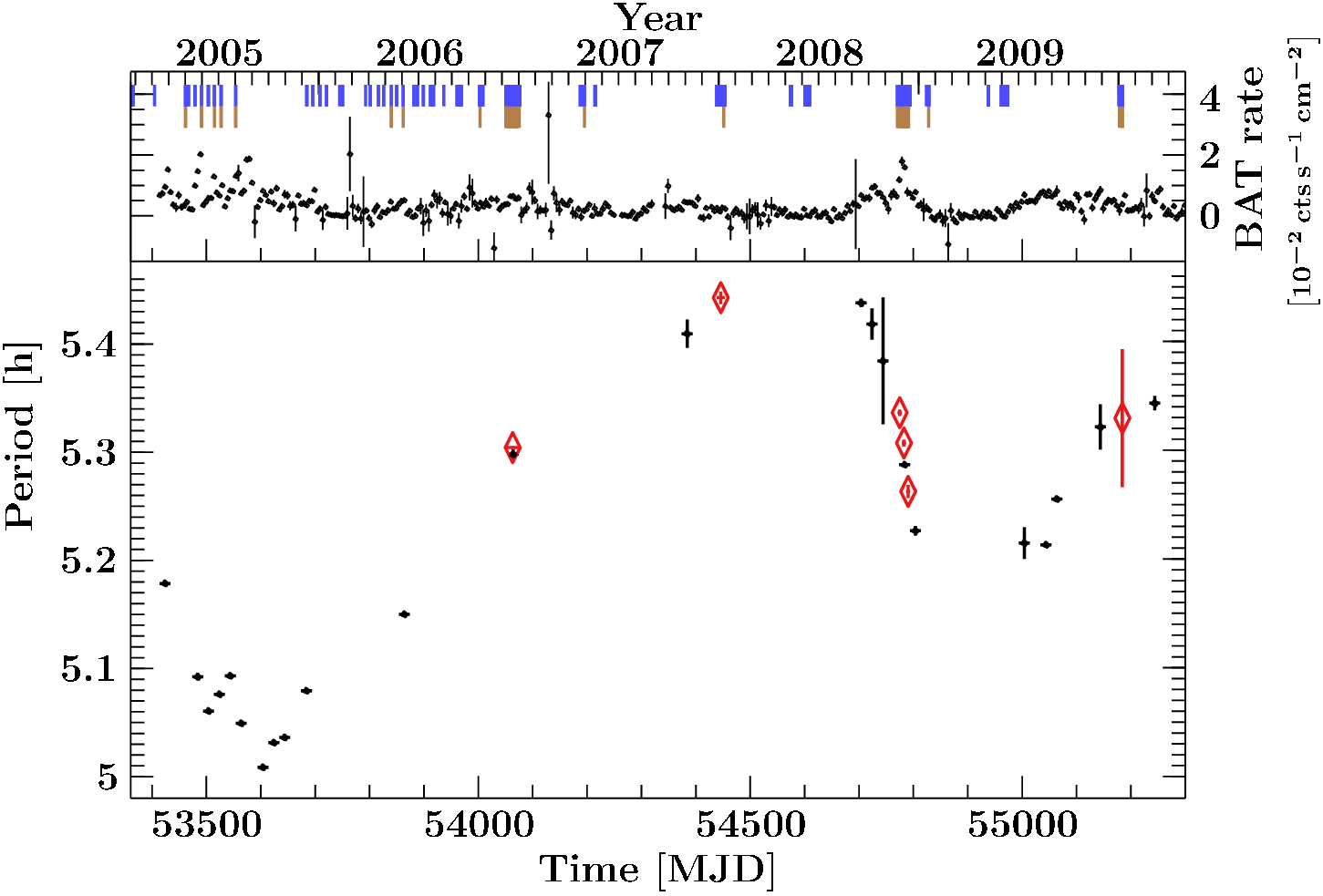}
  \caption{From top to bottom: \textsl{INTEGRAL} observations of
    3A\,1954$+$319 with an offset angle $\leq$$10^\circ$ and
    \textsl{INTEGRAL}-ISGRI detections (blue and brown tickmarks,
    respectively). The long-term light curve shown was obtained by
    \textsl{Swift}-BAT in the 15--50\,keV range and has been rebinned
    to a resolution of 5\,days. The lower part of the figure shows the
    pulse period evolution as determined by BAT (black) and ISGRI
    (red).}\label{fig:evol}
(A color version of this figure is available in the online journal.)
\end{figure}

The upper part of Figure~\ref{fig:evol} shows the 2005--2009
15--50\,keV light curve of 3A\,1954$+$319\footnote{Daily light curve
  from \texttt{http://heasarc.gsfc.nasa.gov/docs/
    swift/results/transients/4U1954p31/}.} observed with
\textsl{Swift}-BAT \citep{barthelmy:05a}.  Months long flaring
episodes are apparent, especially in 2005 and in 2008. The former
includes most of the time range analyzed by \citet{corbet:08a},
$\sim$MJD\,53330--53680. The short tickmarks above the BAT light
curves indicate 1163 \textsl{INTEGRAL} \citep{winkler:03a} pointings
(``science windows'', $\sim$2\,ks exposures) during which
3A\,1954$+$319 was within the field of view of the ISGRI
\citep{lebrun:03a} of the IBIS instrument, with a pointing offset
$\leq$$10^\circ$.

Since its launch in 2002 \textsl{INTEGRAL} has performed several
extensive monitoring campaigns of the Cygnus region
\citep[][]{pottschmidt:03a,cadolle:06a,martin:09a,laurent:11a,williams:11a}.
A Key Program (KP) centered on the black hole X-ray binary Cygnus X-1,
located 3$^\circ$.15 from 3A\,1954$+$319, has been in place since 2008
with annual exposures of several 100\,ks \citep{grinberg:11a}.
Coincidentally these KP observations covered about three weeks of the
2008 flare of 3A\,1954$+$319 in unprecedented detail
(Figure~\ref{fig:evol}).

The 201 science windows for which 3A\,1954$+$319 was detected (OSA
\texttt{DETSIG} $\geq$6 in the 20--100\,keV science window images) are
indicated by long tickmarks in Figure~\ref{fig:evol}. For these
pointings the \texttt{ii\_light} tool in version 7 of the Offline
Scientific Analysis \citep[OSA;][]{courvoisier:03a}\footnote{The
  \texttt{ii\_light} tool was not present in OSA 8 and at the time of
  writing was still undergoing evaluation for OSA 9.} was applied to
produce ISGRI light curves of 3A\,1954$+$319 with a time resolution of
100\,s in the energy bands 20--40\,keV, 40--100\,keV, and
20--100\,keV. Together with the BAT long-term light curve the
20--100\,keV ISGRI light curves were used to determine the pulse
period evolution (Section~\ref{sec:evol}).

In addition, a more detailed analysis was performed for the flare in
2008. This dataset included science windows from satellite revolutions
739, 741--746, 756, and 758 (one revolution takes about three days).
Pulse profiles in the three energy bands were created
(Section~\ref{sec:prof}). In order to maximize the signal to noise
ratio (S/N) all science windows in the phase range of 0.45--0.85,
i.e., associated with the main pulse peak, were selected. OSA 9 was
used to extract average ISGRI and Joint European X-ray Monitor
\citep[JEM-X;][]{lund:03a} spectra, for source offset angles
$\leq$$10^\circ$ and $\leq$$3^\circ$, respectively. The ISGRI spectrum
had an exposure of 85\,ks, was created by averaging spectra from
individual science windows, and was modeled in the 20--80\, keV range.
The JEM-X spectrum had an exposure of 9.8\,ks, was extracted from
mosaic images using \texttt{mosaic\_spec} and was modeled in the
3--30\,keV range (Section~\ref{sec:spec}). Response and auxiliary
response files were selected or created following the analysis
documentation\footnote{See
  http://www.isdc.unige.ch/integral/analysis\#Documentation}.

\section{Results}\label{sec:res}

\subsection{Light Curves}\label{sec:lc}

\begin{figure}
  \includegraphics[width=0.5\textwidth]{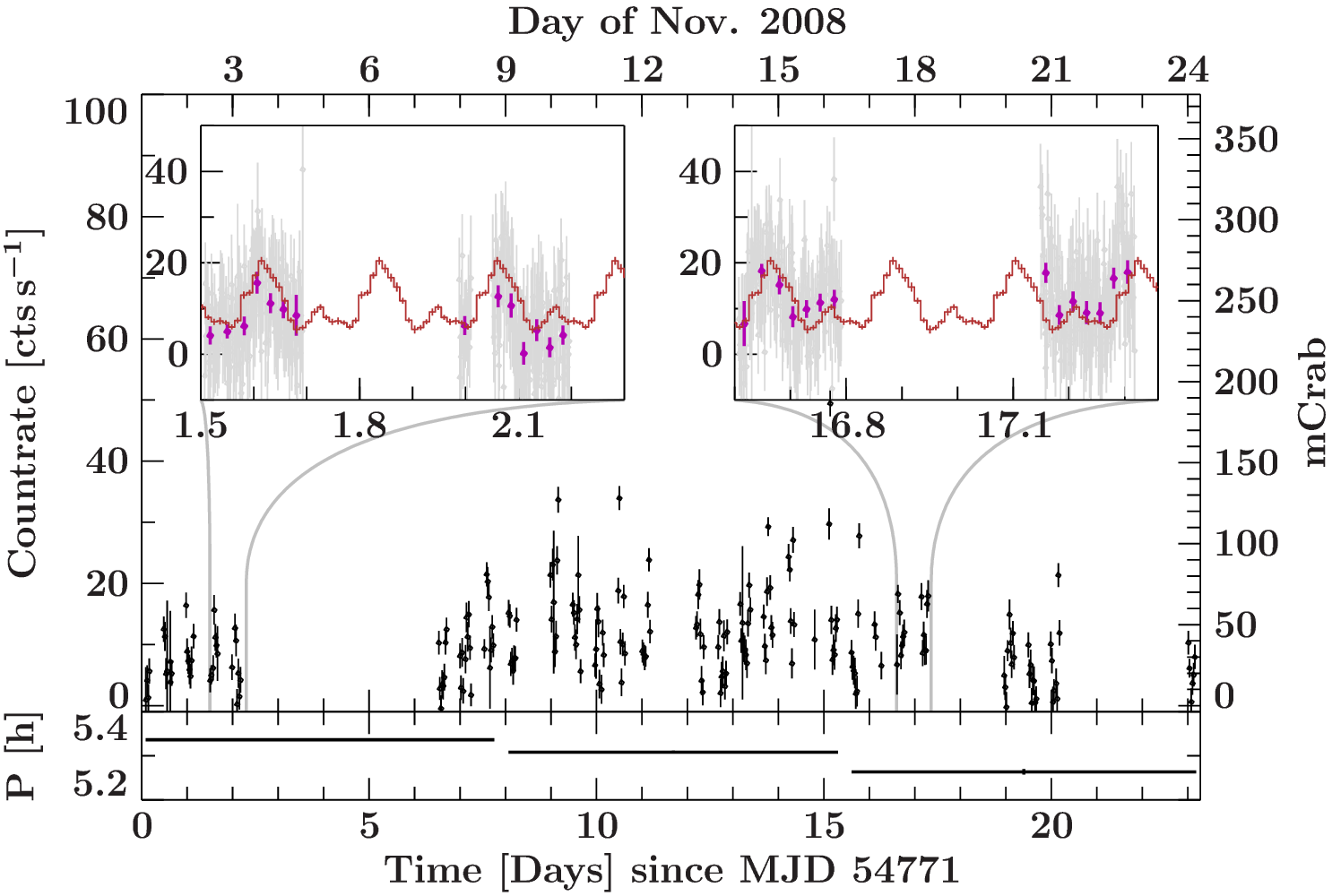}
  \caption{Upper panel: \textsl{INTEGRAL}-ISGRI 20--100\,keV light
    curve of 3A\,1954$+$319 observed during the flaring episode in
    2008. Each data point represents one science window
    ($\sim$2\,ks). The insets show two close-ups with the full 100\,s
    resolution light curve (gray), the count rates per science window
    (magenta), and the average pulse profile (red histogram). Lower
    panel: The three pulse period measurements obtained with ISGRI
    during the flare.}\label{fig:lc}
(A color version of this figure is available in the online journal.)
\end{figure}

The long-term BAT light curve shown in Figure~\ref{fig:evol}
demonstrates the irregular flaring of 3A\,1954$+$319 on timescales of
months, with the 2008 November flare having been one of the brightest
since the start of the BAT monitoring. The \textsl{INTEGRAL}
observations cover the second half of the flaring episode. The
resulting 20--100\,keV ISGRI light curve is shown in the upper panel
of Figure~\ref{fig:lc}. The 20--100\,keV flux during the outburst
varied by a factor of $\sim$20 with an average of $\sim$40\,mCrab and
a peak value of $\sim$130\,mCrab.

The insets of Figure~\ref{fig:lc} show close-ups of two randomly
selected parts of the ISGRI outburst light curve with a resolution of
100\,s. Individual pulses are directly observed in the light curve for
the first time. Also shown are repetitions of the $\dot{P}$-corrected
average pulse profile obtained by folding the high resolution light
curve on the pulse ephemeris determined from the outburst data
(Section~\ref{sec:evol}). Comparing the profiles with the high time
resolution light curve and the average science window count rates
demonstrates general consistency but allows for moderate
pulse-to-pulse variations which are common in accreting X-ray pulsars
\citep{klochkov:11a}.

\subsection{Pulse Profiles}\label{sec:prof}

\begin{figure}
  \includegraphics[width=0.5\textwidth]{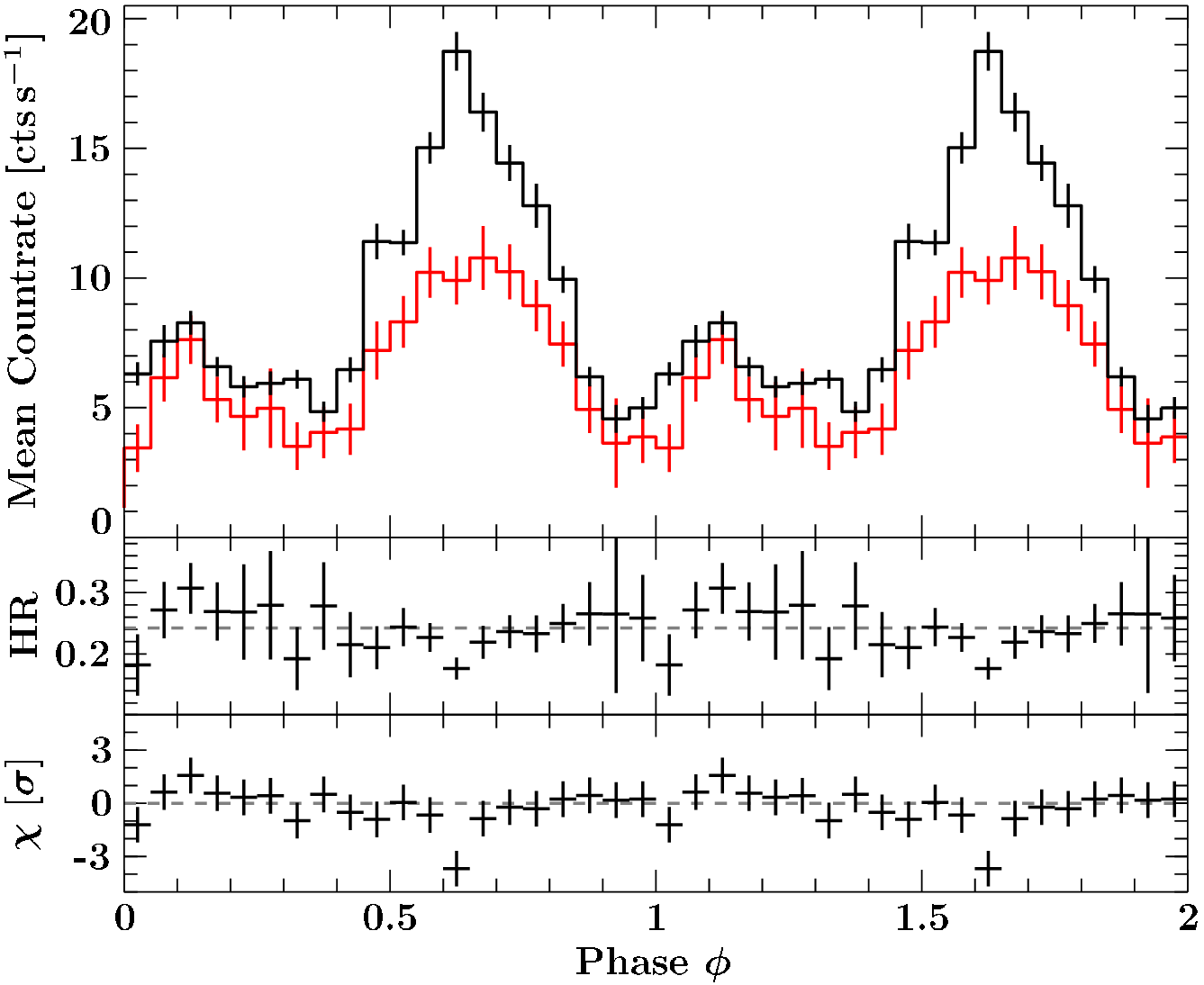}
  \caption{Upper panel: ISGRI pulse profiles for the flaring episode
    in 2008, in the energy ranges of 20--40\,keV (black) and
    40--100\,keV (red, multiplied by 3 for better visibility). Middle
    panel: Hardness ratio obtained by dividing the 40--100\,keV by the
    20--40\,keV profile. The dashed line indicates the mean
    hardness. Lower panel: Deviation of the hardness ratio from the
    mean hardness in units of $\sigma$. The dashed line indicates no
    deviation.}\label{fig:prof}
(A color version of this figure is available in the online journal.)
\end{figure}

Pulse profiles in the energy ranges of 20--40\,keV and 40--100\,keV
were obtained by folding the energy resolved high time resolution
ISGRI outburst light curves on the pulse ephemeris determined from the
2008 outburst data (Section~\ref{sec:evol}), see upper panel of
Figure~\ref{fig:prof}. These are the highest quality pulse profiles
available for the source to date. They clearly show a double peaked
structure, in contrast to the single peaked pulse profile displayed by
the prototype SyXB GX\,1$+$4 \citep{ferrigno:07a}. While the possible
presence of the secondary peak for 3A\,1954$+$319 was indicated in the
$<$50\,keV BAT profiles presented by \citet{corbet:08a}, it could not
be detected in the 50--100\,keV band, possibly due to the
comparatively smaller S/N. The lower two panels of
Figure~\ref{fig:prof} show the hardness ratio between the profiles in
the two energy bands and its deviation from the average ratio. No
significant energy dependence was detected with exception of a
possible moderate softening during the brightest part of the main
pulse in a narrow phase range ($\sim0.60-0.65$).

\subsection{Pulse Period Evolution}\label{sec:evol}

Using the epoch folding technique \citep{schwarzenberg-czerny:89a}
pulse period values were determined for three equally long parts of
the 2008 outburst observations with ISGRI. The results are shown in
the lower panel of Figure~\ref{fig:lc}. Uncertainties were calculated
according to the Monte Carlo method described by \citet{davies:90a}:
for every segment $10^4$ light curves with the same sampling and
variance as seen in the observational data were simulated, based on
the segment's average pulse profile modified by Gaussian noise. A
successful period search was performed for every simulated light curve
and the widths of the emerging distributions were used as
uncertainties of the pulse period measurements. The three ISGRI pulse
period values for the flare evolved from $5.336\pm0.003$\,hr over
$5.308\pm0.003$\,hr to $5.264\pm0.006$\,hr, i.e., a strong spin-up
became apparent. The pulse ephemeris over the 2008 flare was
determinded to $T_0=\mathrm{MJD}54782.6897$,
$P(T_0)=5.3060\pm0.0007$\,hr, and
$\dot{P}(T_0)=(-1.81\pm0.17)\times10^{-4}\,\mathrm{hr}\,\mathrm{hr}^{-1}$
(uncertainties are given on a 1$\sigma$ confidence level).

In order to put the ISGRI period values into perspective the pulse
period history from the long-term BAT light curve was updated by
performing local period determinations for 20\,days long segments. The
lower part of Figure~\ref{fig:evol} contains all successful BAT period
measurements, the three ISGRI flare measurements, as well as three
additional ISGRI measurements that could typically be obtained during
times of denser sampling and/or elevated count
rates\footnote{Following a conservative approach only peaks which were
  visually clearly apparent in the epoch folding statistics were
  selected.}. The long-term results are the following: (i) the spin-up
phase during the flaring activity in 2005, analyzed by
\citet{corbet:08a} and \citet{mattana:06a}, was reproduced, (ii) it
was followed by a long spin-down trend between 2005 and 2008,
characterized by
$\dot{P}\sim2.1\times10^{-5}\,\mathrm{hr}\,\mathrm{hr}^{-1}$, (iii) the
BAT and ISGRI pulse period values are in excellent agreement,
especially during the strong spin-up in 2008, (iv) the spin-down trend
resumed in 2009, possibly slowed down by continued moderate flaring.
Note that while the possibility of a beginning spin-down at the end of
2005 was mentioned by \citet{corbet:08a}, a spin-down has now been
clearly observed for the first time for 3A\,1954$+$319. Also note that
the spin-up in 2008 was an order of magnitude larger than the one in
2005.

\subsection{Broadband Spectrum}\label{sec:spec}

\begin{figure}
  \includegraphics[width=0.5\textwidth]{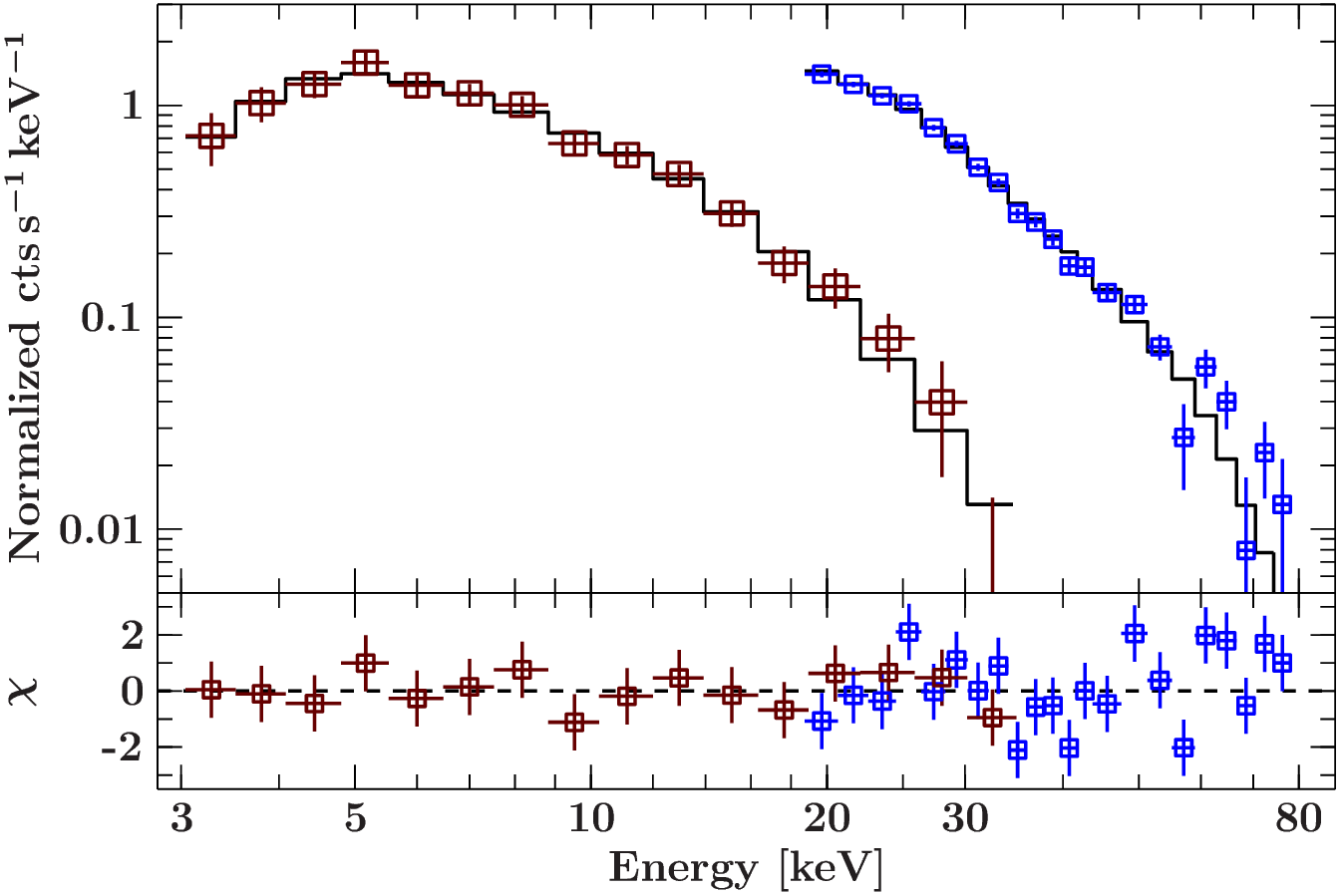}
  \caption{Upper panel: JEM-X (3--30\,keV, brown) and ISGRI
    (20--80\,keV, blue) counts spectra for the flaring episode in
    2008. In order to obtain a better S/N only data from the main
    pulse, i.e., phases 0.45--0.85, were included. The best fit
    description consisted of an absorbed thermal Comptonization model
    and is also displayed. Lower panel: Best fit
    residuals.}\label{fig:spec}
(A color version of this figure is available in the online journal.)
\end{figure}

As described above the 3--80\,keV spectrum of 3A\,1954$+$319 was
determined for the phase range associated with the peak of the main
pulse. This approach was chosen because the source is close to the
detection limit during dimmer pulse phases. The spectrum could be well
described by an absorbed thermal Comptonizaton model
(Figure~\ref{fig:spec}), resulting in $\chi^2_\mathrm{red}=1.3$ for
32\,degrees of freedom. The best fit parameters obtained were
$N_\mathrm{H}=3.9^{+6.7}_{-2.2}\times10^{22}\,\mathrm{cm}^{-2}$,
$kT_0=0.9^{+0.4}_{-0.9}$\,keV, $\tau=4.2\pm0.7$, and
$kT_\mathrm{e}=7.5^{+0.5}_{-0.4}$\,keV. A flux cross calibration
factor of $0.84^{+0.12}_{-0.11}$ was applied to model the JEM-X data.
The residuals mainly reflect known ISGRI calibration uncertainties
\citep{grinberg:11a}. All fit parameter uncertainties, including those
of the flux measurements below, are given at the 90\% level for one
interesting parameter. A broken power law fit provided a slightly
better description but the improvement is due to masking
cross-calibration issues ($\chi^2\sim1.2$). A cutoff power law model
as applied by \citet{mattana:06a} resulted in a worse fit
($\chi^2\sim1.9$).

The absorbed 2--10\,keV and 10--100\,keV fluxes were
$7.2^{+0.8}_{-0.7}$ and
$12.1^{+0.3}_{-0.3}\times10^{-10}\,\mathrm{erg}\,\mathrm{cm}^{-2}\,\mathrm{s}^{-1}$,
respectively (ISGRI normalization). The unabsorbed fluxes were
9.1$^{+2.7}_{-2.0}$ and
$12.2^{+0.4}_{-0.3}\times10^{-10}\,\mathrm{erg}\,\mathrm{cm}^{-2}\,\mathrm{s}^{-1}$,
corresponding to (pulse peak) luminosities of $3.1$ and
$4.2\times10^{35}\,\mathrm{erg}\,\mathrm{s}^{-1}$ for a distance of
1.7\,kpc. Since the assumed distance is an upper limit
\citep{masetti:06a}, these luminosities are upper limits as well. They
are comparable to the highest luminosities reported by
\citet{masetti:07a}.

\section{Discussion}\label{sec:dis}

A possible scenario for the development of pulse periods longer than a
few hundred seconds has been proposed for persistently bright HMXBs by
\citet{ikhsanov:07a}, where accretion proceeds spherically and a
strong spin-down happened in a previous accretion epoch (subsonic
propeller regime). The author argues that the 2.7\,hr pulse period of
the HMXB 2S\,0114$+$650 \citep{farrell:08a} can thus be explained
without the need for an unusually high ($10^{15}$\,G) magnetic field.

For the LMXB 3A\,1954$+$319 the long pulse period is more easily
reconcilable with the system's life time
\citep{mattana:06a}. \citet{shakura:11a} recently developed an
accretion model for SyXBs that can not only account for long spin
periods but also provides a mechanism allowing for quasi-spherical
accretion. In this model a subsonic settling regime occurs for X-ray
luminosities below
$\sim$$3\times10^{36}\,\mathrm{erg}\,\mathrm{s}^{-1}$, i.e., for
luminosities consistent with those observed for 3A\,1954$+$319. A
shell of hot material forms around the magnetosphere which mediates
the transfer of angular momentum to/from the neutron star by advection
and viscous stress. The accretion rate is determined by the ability of
the plasma to enter the magnetosphere. The equilibrium pulse period in
this case is:
\begin{equation}
\frac{P_\mathrm{eq}}{10^3\,\mathrm{s}} \sim
\left(\frac{B}{10^{12}\,\mathrm{G}}\right)^{12/11}
\left(\frac{P_\mathrm{orbit}}{10\,\mathrm{d}}\right)
\left(\frac{L_\mathrm{X}}{10^{36}\,\mathrm{erg}\,\mathrm{s}^{-1}}\right)^{4/11}
\left(\frac{v_\mathrm{wind}}{10^3\,\mathrm{km}\,\mathrm{s}^{-1}}\right)^{4}
\end{equation} 
where $P_\text{orbit}$ is the binary period and $v_\mathrm{wind}$ is
the stellar wind velocity with respect to the neutron star. Reporting
on population synthesis simulations for SyXBs based on this model,
\citet{postnov:11a} showed that the 5.3\,hr pulse period of
3A\,1954$+$319 can be well reproduced.

3A\,1954$+$319 seems to show long spin-down episodes between major
flares ($\dot{P}\sim2.1\times10^{-5}\,\mathrm{hr}\,\mathrm{hr}^{-1}$)
which are reminiscent of the spin-down displayed by GX\,1$+$4 since
the early 1980s
\citep[$\dot{P}\sim10^{-7}\,\mathrm{hr}\,\mathrm{hr}^{-1}$;][]{gonzalez:11a}.
Several models beyond equilibrium disk accretion have been proposed to
explain the spin-down of GX\,1$+$4 without invoking unusually high
magnetic fields, for example the presence of a counterrotating disk
\citep{nelson:97a} or accretion of fallback material expelled during
the propeller phase \citep{perna:06a}. Only the quasi-spherical
accretion model, however, reproduces the correct sign and magnitude of
the negative correlation between spin frequency change $\dot{\nu}$ and
X-ray flux $F_\mathrm{X}$ observed during spin-down in this source
\citep[$-\dot{\nu}\propto F_\mathrm{X}^{3/7}$;][]{gonzalez:11a}. For
3A\,1954$+$319 such a detailed study of the
$\dot{\nu}$-$F_X$-relationship is difficult, especially during
spin-down since the low flux allowed for only a few $\dot{\nu}$
measurements (Figure~\ref{fig:evol}). It is beyond the scope of this
Letter.

On longer time scales we observe torque reversals and a positive
$\dot{\nu}$-$F_\mathrm{X}$-correlation between low (spin-down) and
high flux (spin-up) episodes (Figure~\ref{fig:evol}).  This is
consistent with the behavior predicted by the quasi-spherical
accretion model for higher accretion rates within the settling regime
\citep[see Figure~1 of][]{shakura:11a}.

The strong spin-up in 2008 translates to
$\dot{P}/P=-0.9\times10^{-8}\,\mathrm{s}^{-1}$. While still high, the
absolute value is of the same order of magnitude as the spin-down
related $\dot{P}/P$ of $3.1\times10^{-8}\,\mathrm{s}^{-1}$ observed
for the SyXB IGR\,J16358$-$4724 \citep{patel:07a}. The X-ray
luminosity required to sustain such a spin-up in 3A\,1954$+$319 in the
equilibrium disk accretion case would be
$\sim$$5\times10^{36}\,\mathrm{erg}\,\mathrm{s}^{-1}$
\citep{joss:84a}, whereas we obtained an upper limit for the pulse
peak flux value of
$\sim$$7.4\times10^{35}\,\mathrm{erg}\,\mathrm{s}^{-1}$, again arguing
against disk accretion.

This conclusion does not change when the possible contribution of a
spin-up due to orbital motion is considered: According to
\citet{dumm:98a} the measured maximum mass for M giants is
$\sim3.5$\,$M_\sun$ and the median stellar radii determined for M4~III
and M5~III stars are 103\,$R_\sun$ and 120\,$R_\sun$,
respectively. For neutron star orbits outside of the M giant, the
orbital period then has to be $\gtrsim80$\,days and
$\gtrsim100$\,days, translating into a maximum relative change of the
pulse period over the orbit of $\lesssim4\times10^{-4}$ in both
cases. The lowest relative uncertainty range of our pulse period
measurements is $\sim10^{-3}$, considerably bigger than any realistic
orbital effect, which will be even smaller than the value above due to
a typically lower stellar mass and wider orbit. Therefore the orbital
influence is negligible for the 2008 spin-up measurement.

The best fit to the broadband spectrum of the 2008 flare describes an
optically thick Compton plasma ($\tau=4.2\pm0.7$,
$kT_\text{e}=7.5^{+0.5}_{-0.4}$\,keV) with parameters qualitatively
consistent with the results of \citet{masetti:07a}. As these authors
state the parameters are similar to those commonly seen in LMXBs with
an accreting neutron star companion. Most of those sources do not show
pulsations, however, and their accretion geometry is most likely
different from that of pulsars. Other SyXBs show similar spectra,
especially GX\,1$+$4 \citep[$\tau=6.80\pm0.15$,
  $kT_\text{e}=13.1\pm0.2$\,keV;][]{ferrigno:07a}. Furthermore,
optically thick Comptonization has also been used to describe the
emission from the accretion columns of accreting pulsars in HMXBs,
e.g., for the cyclotron line soure 1A\,1118$-$61 \citep{suchy:11a}. We
tentatively propose a similar origin close to the neutron star surface
for the broadband X-ray emission of 3A\,1954$+$319.

\acknowledgments

We thank the anonymous referee for useful comments. DMM and KP
acknowledge NASA grants NNX08AE84G, NNX08AY24G, and NNX09AT28G. FF
acknowledges support from the DAAD and thanks the NASA-GSFC for its
hospitality. The work by KAP is partially supported through RFBR grant
10-02-00599. This research has been partly funded by the European
Commission under contract ITN215212 ``Black Hole Universe'' and by the
Bundesministerium f\"ur Wirtschaft and Technologie under DLR grants
50OR0808 and 50OR1007. It is based on observations with
\textsl{INTEGRAL}, an ESA project with instruments and science data
centre funded by ESA member states (especially the PI countries:
Denmark, France, Germany, Italy, Switzerland, Spain), Czech Republic
and Poland, and with the participation of Russia and the USA.  We
thank the \textsl{INTEGRAL} mission planners for careful scheduling of
the Cygnus region Key Program. We also thank Hans Krimm and the
\textsl{Swift}-BAT team for making the \textsl{Swift}-BAT light curves
available.

Facilities: \facility{INTEGRAL}, \facility{Swift}.

\end{document}